\documentclass[a4paper]{article}
\usepackage{graphicx}
\usepackage{amsfonts}
\usepackage{amsmath}
\usepackage{amssymb}
\DeclareMathOperator{\Tr}{Tr}

\begin{document}

\title{Contact potential instability in the path-integral description of itinerant ferromagnetism}

\author{E. Vermeyen and J. Tempere\\\textit{TQC (Theory of Quantum systems and Complex systems)}\\\textit{Universiteit Antwerpen, 2610 Wilrijk, Belgium}}

\date{15th December 2014}

\maketitle

\begin{abstract}
It has long been predicted that a two-component non-localized Fermi gas will exhibit spontaneous polarization for sufficiently strong repulsive interactions, a phenomenon which is called itinerant ferromagnetism.  Recent experiments with ultracold atomic gases have reached the interaction strength for which theoretical models have predicted the occurrence of the normal-to-itinerant-ferromagnetic phase transition, but so far this transition has not been observed. The instability of the repulsive branch of the Feshbach resonance prevents the formation of the itinerant ferromagnetic state, but it is not clear whether this is the only instability impeding its experimental realization. In this article, we use the path-integral formalism with density fields in the Hubbard-Stratonovich transformation to study the stability of a homogeneous two-component Fermi gas with contact interactions. Within the saddle-point approximation we show that none of the extrema of the action are minima, meaning all extrema are unstable to small density fluctuations. This implies a more general mechanical instability of the polarized (itinerant ferromagnetic) and normal states of the system in the path-integral formalism.  We find that it is important to consider the stability of the system when studying itinerant ferromagnetism. Since (mechanical) stability may be influenced by the details of the interaction potential, we suggest the use of a more realistic potential than the contact potential in future theoretical descriptions.
\end{abstract}

\section{Introduction}

Itinerant ferromagnetism is defined as spontaneous polarization of non-loca\-lized particles. In two-component Fermi gases, we can expect it to occur for strongly repulsive interactions: due to the exchange interaction, polarization reduces the interaction energy of the gas, at the cost of an increased kinetic energy. It was originally predicted by Bloch in 1929 \cite{Bloch} for electrons and the theoretical groundwork was refined by Stoner \cite{Stoner} shortly thereafter. Despite its long history, itinerant ferromagnetism is not yet well understood. The strong correlations involved make it difficult to create an accurate theoretical description and it is hard to find a good experimental system in which it can be studied. In solid-state systems it occurs in d-band transition metals (e.g. Fe, Ni, Co), but the influence of the exact shape of the band and the mixture of both localized and itinerant ferromagnetism make it hard to verify models for itinerant ferromagnetism in these systems \cite{dband}. Ultracold atomic gases have already proven to be a fertile testing ground for several theoretical models (e.g. the BCS-BEC crossover \cite{bcsbec1,bcsbec2,bcsbec3,bcsbec4,bcsbec5,bcsbectheory1,bcsbectheory2}). Spin-polarization has been investigated in the superfluid state, both experimentally \cite{spinsusceptibility1} and theoretically \cite{spinsusceptibility2}. Several studies have proposed that, also in the normal state, quantum gases could serve as a model system for itinerant ferromagnetism \cite{ifprediction1,ifprediction2,ifprediction3,ifprediction4,RADuineprediction}.

In 2009, some tantalizing experimental hints of the normal-to-itinerant-ferro\-magnetic phase transition were observed \cite{MITexp}, but the final confirmation of the nature of the observed transition was missing: no magnetic domains could be observed. Interestingly, the observed transition was found at the dimensionless scattering length $k_{F}a_{s}=1.9 \pm 0.2$, a value which is significantly larger than the mean-field prediction $k_{F}a_{s}=\pi/2$ or the second order prediction $k_{F}a_{s}=1.054$ \cite{RADuineprediction}. These results renewed the theoretical interest in itinerant ferromagnetism \cite{GJConduitpadintegralen,followupstudy1,followupstudy2,followupstudy3,followupstudy4,followupstudy5}. In a subsequent experiment \cite{opvolgingsexperiment} it was found that the formation of the itinerant ferromagnetic state is prevented by a rapid decay into bound pairs due to near-resonant three-body interactions. Experimentalists are currently looking for a way to suppress this decay; the most promising proposal is the use of a mixture of $^{40}$K and $^{6}$Li \cite{massimbalance}. There are also several proposals that try to avoid the decay by reducing the critical interaction strength of the normal-to-itinerant-ferromagnetic phase transition \cite{latticeitferro,socitferro}. However, it is not clear if this instability of the positive branch of the Feshbach branch resonance to molecular pairing is the only instability preventing the formation of the itinerant ferromagnetic state.

In this paper we revisit the itinerant ferromagnetic state using the path-integral formalism, with the goal of studying its stability with respect to small (density) fluctuations within the saddle-point approximation. We start from a homogeneous two-component Fermi gas with contact interactions. The partition function corresponding to this system, written as a path integral, cannot be calculated exactly for a general case. A Hubbard-Stratonovich transformation is used to rewrite the fermionic path integral into a form that can be calculated exactly, at the cost of introducing an auxiliary bosonic field. In the saddle-point approximation we assume the auxiliary bosonic field to be constant and we require its value to be a minimum, maximum or saddle point (henceforth called extremum) of the action. In order to study itinerant ferromagnetism, the auxiliary bosonic field is chosen to be a density field (also called the Hartree channel). An extremum of the action is stable to small fluctuations if it is also a minimum. In order to study their stability, we study the nature of all extrema. We find that all extrema of the action for the density fields are unstable against small (density) fluctuations and we discuss the possible implications of this instability.

This paper is organized as follows: In Sec. \ref{treatment}, we construct the path-integral treatment of the problem, including a detailed description of the Hubbard-Stratonovich transformation. In Sec. \ref{saddlepoint}, we introduce the saddle-point approximation and we look into the details of the resulting expression for the saddle-point thermodynamic grand potential per unit volume. In Sec. \ref{stability}, the stability of the extrema of the action are studied and it is shown that none of the extrema are minima. In Sec. \ref{conclusiondiscussion}, we discuss the results and their possible implications.

\section{The path-integral treatment\label{treatment}}

The thermodynamic grand potential $\Omega$ per unit volume can be determined from the grand-canonical partition sum $Z$,%
\begin{equation}
\Omega=-\frac{1}{\beta V}\ln Z
\end{equation}
with $\beta=1/k_{B}T$ the inverse temperature, $T$ the temperature, $k_{B}$ the Boltzmann constant and $V$ the volume. In the path-integral formalism, the grand-canonical partition sum of a two-component Fermi gas is given by%
\begin{equation}
Z=\prod_{\sigma=\uparrow,\downarrow}\left(  \int\mathcal{D}\bar{\psi}_{\sigma}\int\mathcal{D}\psi_{\sigma}\right)  \exp\left(  -S\left[  \bar{\psi}_{\sigma},\psi_{\sigma}\right]  \right)  \text{,} \label{sumofstates}
\end{equation}
where we take the sum over all possible configurations of the Grassmann fields $\bar{\psi}_{\uparrow}$, $\psi_{\uparrow}$, %
$\bar{\psi}_{\downarrow}$ and $\psi_{\downarrow}$, weighted by a factor that depends on the action $S$ of each specific configuration. The two components are called spin-up $\uparrow$ and spin-down $\downarrow$, for example referring to two different hyperfine states of an atom (in the context of ultracold Fermi gases). For a uniform Fermi gas with contact interactions, the action (in units $\hbar=1$, $2m=1$ and $k_B=1$) is given by%
\begin{align}
S\left[  \bar{\psi}_{\sigma},\psi_{\sigma}\right] = & \sum_{\sigma=\uparrow,\downarrow}\int_{0}^{\beta}d\tau\int d\mathbf{x}\bar{\psi
}_{\mathbf{x}\tau\sigma}\left[  \frac{\partial}{\partial\tau}-\mathbf{\nabla
}_{\mathbf{x}}^{2}-\mu_{\sigma}\right]  \psi_{\mathbf{x}\tau\sigma}\nonumber\\
& +g\int_{0}^{\beta}d\tau\int d\mathbf{x}\bar{\psi}_{\mathbf{x}\tau\uparrow
}\bar{\psi}_{\mathbf{x}\tau\downarrow}\psi_{\mathbf{x}\tau\downarrow}%
\psi_{\mathbf{x}\tau\uparrow}\text{,} \label{action}%
\end{align}
with the Grassmann fields now expressed as a function of imaginary time $\tau$ and position vector $\mathbf{x}$. In (\ref{action}), $\mu_{\sigma}$ is the chemical potential of particles in spin state $\sigma$ and $g$ is the strength of the interaction potential, in 3D related to the s-wave scattering length $a_{s}$ by $g=4\pi\hbar^{2}a_{s}/m$ (or $g=8\pi a_{s}$ in the chosen units). Due to the presence of the interaction term in (\ref{action}), which is of fourth order in the fermionic fields, the path integral in (\ref{sumofstates}) cannot be solved exactly for a general case (although an exact solution is available for the 1D case with contact interactions \cite{1Dexactsolution}). The Hubbard-Stratonovich transformation is used to transform the interaction term into a form which is quadratic in the fermionic fields (and thus can be calculated exactly), at the cost of introducing an extra path integral over an auxiliary bosonic field. This way, the problem is shifted to the bosonic path integral, which usually has to be approximated.

There are many different ways to decouple the quartic interaction term (using a Hubbard-Stratonovich transformation). They fall into three categories, referring to the three possible ways to divide the four fermionic fields of the interaction term into two pairs \cite{KleinertHS}:%
\begin{itemize}
\item Bogoliubov: $\bar{\psi}_{\mathbf{x}\tau\uparrow}\bar{\psi}_{\mathbf{x}\tau\downarrow}$ and $\psi_{\mathbf{x}\tau\downarrow}\psi_{\mathbf{x}\tau\uparrow}$
\item Fock: $\bar{\psi}_{\mathbf{x}\tau\uparrow}\psi_{\mathbf{x}\tau\downarrow}$ and $\bar{\psi}_{\mathbf{x}\tau\downarrow}\psi
_{\mathbf{x}\tau\uparrow}$
\item Hartree: $\bar{\psi}_{\mathbf{x}\tau\uparrow}\psi_{\mathbf{x}\tau\uparrow}$ and $\bar{\psi}_{\mathbf{x}\tau\downarrow}\psi_{\mathbf{x}\tau\downarrow}$
\end{itemize}
The Bogoliubov channel is suitable to describe superfluidity and focusses on fer\-mion\-ic pair formation \cite{stoofboek,pairingfields1,pairingfields2,pairingfields3}. The Fock channel represents spin-flip interactions. In this paper we use the Hartree channel, as the density fields capture the essence of the interactions in the normal and polarized states \cite{densityfields1,densityfields2,densityfields3}. The Hubbard-Stratonovich transformation itself is exact for any channel, but the choice of the channel determines the physics that is included in the saddle-point approximation for the corresponding auxiliary bosonic field. That is why the choice of the decoupling of the interaction term is a very important step in our description.

In the Hartree channel still two kinds of density fields appear, corresponding to the density in the two different spin states. Together with their conjugated counterparts, we end up with four density fields: $\bar{\rho}_{\uparrow}$, $\rho_{\uparrow}$, $\bar{\rho}_{\downarrow}$ and $\rho_{\downarrow}$. For symmetry reasons two separate transformations are used, each corresponding to one half of the interaction energy:%

\begin{figure}
[ptb]
\includegraphics[scale=0.72]{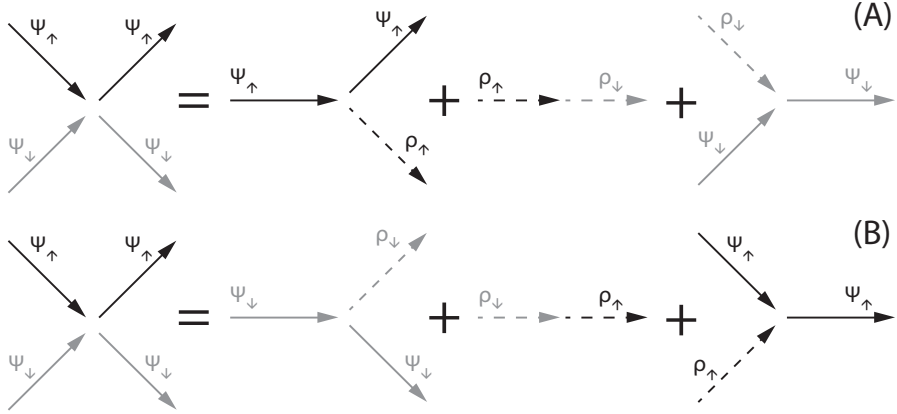}%
\caption{The Feynman diagrams corresponding to the two Hubbard-Stratonovich transformations: (A) corresponds to (\ref{HStransform1}) and (B) corresponds to (\ref{HStransform2}). The black arrows represent spin-up fields, the grey arrows represent spin-down fields. The bosonic fields $\rho_{\uparrow}$ and $\rho_{\downarrow}$ (dashed arrows) are introduced as a mediator of the interaction between the fermionic fields (full arrows). The right hand sides of (A) and (B) show the same process in opposite directions, so both transformations are complementary.}%
\label{diagrammen} 
\end{figure}

\begin{align}
&  \exp\left(  -\frac{g}{2}\int_{0}^{\beta}d\tau\int d\mathbf{x}\bar{\psi
}_{\mathbf{x}\tau\uparrow}\bar{\psi}_{\mathbf{x}\tau\downarrow}\psi
_{\mathbf{x}\tau\downarrow}\psi_{\mathbf{x}\tau\uparrow}\right) =\label{HStransform1}\\
&\int\mathcal{D}\bar{\rho}_{\downarrow}\int\mathcal{D}%
\rho_{\uparrow}\exp\left[ \frac{1}{2} \int_{0}^{\beta}d\tau\int d\mathbf{x}\left(
-\bar{\rho}_{\mathbf{x}\tau\downarrow}\bar{\psi}_{\mathbf{x}\tau\downarrow
}\psi_{\mathbf{x}\tau\downarrow}+\frac{\bar{\rho}_{\mathbf{x}\tau\downarrow
}\rho_{\mathbf{x}\tau\uparrow}}{g}-\bar{\psi}_{\mathbf{x}\tau\uparrow}%
\psi_{\mathbf{x}\tau\uparrow}\rho_{\mathbf{x}\tau\uparrow}\right)  \right]\text{,}
\nonumber
\end{align}
\begin{align}
&  \exp\left(  -\frac{g}{2}\int_{0}^{\beta}d\tau\int d\mathbf{x}\bar{\psi
}_{\mathbf{x}\tau\uparrow}\bar{\psi}_{\mathbf{x}\tau\downarrow}\psi
_{\mathbf{x}\tau\downarrow}\psi_{\mathbf{x}\tau\uparrow}\right) =\label{HStransform2}\\
& \int\mathcal{D}\bar{\rho}_{\uparrow}\int\mathcal{D}%
\rho_{\downarrow}\exp\left[ \frac{1}{2} \int_{0}^{\beta}d\tau\int d\mathbf{x}\left(
-\bar{\rho}_{\mathbf{x}\tau\uparrow}\bar{\psi}_{\mathbf{x}\tau\uparrow}%
\psi_{\mathbf{x}\tau\uparrow}+\frac{\bar{\rho}_{\mathbf{x}\tau\uparrow}%
\rho_{\mathbf{x}\tau\downarrow}}{g}-\bar{\psi}_{\mathbf{x}\tau\downarrow}%
\psi_{\mathbf{x}\tau\downarrow}\rho_{\mathbf{x}\tau\downarrow}\right)
\right]  \text{.}\nonumber
\end{align}

These transformations can be viewed as two directions of the same interaction process (Fig. \ref{diagrammen}). After the transformation, the partition sum is given by%
\begin{align}
Z= & \prod_{\sigma=\uparrow,\downarrow}\left(  \int\mathcal{D}\bar{\psi}_{\sigma
}\int\mathcal{D}\psi_{\sigma}\int\mathcal{D}\bar{\rho}_{\sigma}\int%
\mathcal{D}\rho_{\sigma}\right) \exp\left(  -S\left[  \bar{\psi}_{\sigma
},\psi_{\sigma},\bar{\rho}_{\sigma},\rho_{\sigma}\right]  \right)
\end{align}
with%
\begin{align}
& S\left[  \bar{\psi}_{\sigma},\psi_{\sigma},\bar{\rho}_{\sigma},\rho_{\sigma
}\right] = \nonumber\\
& \sum_{\sigma=\uparrow,\downarrow}\int_{0}^{\beta}d\tau\int d\mathbf{x}
\left[  \bar{\psi}_{\mathbf{x}\tau\sigma}\left(  \frac{\partial
}{\partial\tau}-\mathbf{\nabla}_{\mathbf{x}}^{2}-\mu_{\sigma}+\frac{\bar{\rho
}_{\mathbf{x}\tau\left(-\sigma\right)}+\rho_{\mathbf{x}\tau\left(-\sigma\right)}}{2}\right)  \psi
_{\mathbf{x}\tau\sigma}\right] \nonumber\\
&  -\frac{1}{2}\int_{0}^{\beta}d\tau\int d\mathbf{x}\left(  \frac{\bar{\rho}%
_{\mathbf{x}\tau\downarrow}\rho_{\mathbf{x}\tau\uparrow}}{g}+\frac{\bar{\rho
}_{\mathbf{x}\tau\uparrow}\rho_{\mathbf{x}\tau\downarrow}}{g}\right) \text{.}\label{ZafterHS}%
\end{align}

The derivatives in (\ref{ZafterHS}) are removed by a Fourier transformation. Subsequently we change to the Nambu spinor notation and finally the fermionic path integral is performed. Together with a change in notation for the density fields and the chemical potentials,%
\begin{equation}
\left\{
\begin{array}
[c]{c}%
\rho=\dfrac{\rho_{\uparrow}+\rho_{\downarrow}}{2}\\
\phi=\dfrac{\rho_{\uparrow}-\rho_{\downarrow}}{2}%
\end{array}
\right.  \text{ and }\left\{
\begin{array}
[c]{c}%
\mu=\dfrac{\mu_{\uparrow}+\mu_{\downarrow}}{2}\\
\zeta=\dfrac{\mu_{\uparrow}-\mu_{\downarrow}}{2}%
\end{array}
\right. \text{,}
\end{equation}
corresponding to the average density (resp. chemical potential) and half the density (resp. chemical potential) difference between the spin states, this results in%
\begin{align}
& Z =\prod_{\sigma=\uparrow,\downarrow}\left(  \int\mathcal{D}\bar{\rho
}_{\sigma}\int\mathcal{D}\rho_{\sigma}\right) 
 \exp\left(  \sum_{k}\frac{\left(  \bar{\rho}_{k}\rho_{k}-\bar{\phi}_{k}%
\phi_{k}\right)  }{g}+\operatorname*{Tr}\left\{  \ln\left[  -\det_{\sigma
}\left(  -\mathbb{G}_{k;k^{\prime}}^{-1}\right)  \right]  \right\}  \right)
\text{.}\label{Zafterfermion}
\end{align}
In this expression, we used the short notation $k=\left(  \mathbf{k}%
,\omega_{n}\right)  $ with $\mathbf{k}$ the wavevector and $\omega_{n}=\left(  2n+1\right)  \pi/\beta$ the
fermionic Matsubara frequencies ($n\in\mathbb{Z}$). $-\mathbb{G}^{-1}$ is a matrix in $k$-space, but each matrix element is in itself a $2\times2$ matrix over the spin states:%
\begin{align}
&  -\mathbb{G}_{k,k^{\prime}}^{-1}\left.  =\right.
\begin{pmatrix}
-i\omega_{n}+\mathbf{k}^{2}-\mu-\zeta & 0\\
0 & -i\omega_{n}-\mathbf{k}^{2}+\mu-\zeta%
\end{pmatrix}
\delta\left(  \Delta k\right) \label{Gmatrix}\\
&  +\frac{1}{2\sqrt{\beta V}}%
\begin{pmatrix}
\left(  \bar{\rho}_{-\Delta k}+\rho_{\Delta k}\right)  -\left(  \bar{\phi
}_{-\Delta k}+\phi_{\Delta k}\right)  & 0\\
0 & \left(  \bar{\rho}_{\Delta k}+\rho_{-\Delta k}\right)  +\left(  \bar
{\phi}_{\Delta k}+\phi_{-\Delta k}\right)
\end{pmatrix}
\nonumber
\end{align}
with $\Delta k=k-k^{\prime}$ and $\delta\left(  \Delta k\right)$ the Dirac delta function. The determinant $\det_{\sigma}$ in (\ref{Zafterfermion}) works on the $2\times2$ matrices that are the matrix elements of $-\mathbb{G}^{-1}$, while the trace in (\ref{Zafterfermion}) is taken over $k$-space.

\section{The saddle-point approximation\label{saddlepoint}}

The matrix $-\mathbb{G}^{-1}$ is non-diagonal in $k$-space. Because of this, the bosonic path integral cannot be solved exactly, necessitating approximations. Without interactions, only the first term of (\ref{Gmatrix}) would remain, which is diagonal in $k$. The saddle-point approximation makes the second term in (\ref{Gmatrix}) diagonal in $k$, by assuming only the $\Delta k=0$ terms are important,%
\begin{equation}
\left\{
\begin{array}{l}%
\rho_{\Delta k}=\sqrt{\beta V}\delta\left(\Delta k\right)\rho\\
\bar{\rho}_{\Delta k}=\sqrt{\beta V}\delta\left(  \Delta k\right)\rho^{\ast}%
\end{array}
\right.  \text{,} \label{spcondition1}%
\end{equation}%
\begin{equation}
\left\{
\begin{array}{l}%
\phi_{\Delta k}=\sqrt{\beta V}\delta\left(\Delta k\right)\phi\\
\bar{\phi}_{\Delta k}=\sqrt{\beta V}\delta\left(\Delta k\right)\phi^{\ast}%
\end{array}
\right.  \text{.} \label{spcondition2}%
\end{equation}
The average density and half the density difference are assumed to be constant, hence we neglect all density fluctuations. The saddle-point thermodynamic grand potential per unit volume then becomes
\begin{align}
& \Omega_{sp}\left(  \beta,\mu,\zeta;\rho,\phi\right)=\frac{1}{g}\left(
\phi_{I}^{2}-\rho_{I}^{2}\right)  +\frac{1}{g}\left(  \phi_{R}^{2}-\rho
_{R}^{2}\right) \nonumber \\
&  -\frac{1}{\beta V}\sum_{\mathbf{k},n}\ln\left[  -\left(  -i\omega
_{n}+E_{\mathbf{k}}-\zeta^{\prime}\right)  \left(  -i\omega_{n}-E_{\mathbf{k}%
}-\zeta^{\prime}\right)  \right]  \text{.}\label{Wsp}
\end{align}
Here, we used the following notations to shorten the expressions. The indices $R$ and $I$ refer to the real and imaginary part of $\rho$ and $\phi$. $E_{\mathbf{k}}=\mathbf{k}^{2}-\mu^{\prime}$ is the single-particle dispersion relation. $\mu^{\prime}=\mu-\rho_{R}$ and $\zeta^{\prime}=\zeta+\phi_{R}$ are the effective chemical potentials. We now see that the imaginary parts $\rho_{I}$ and $\phi_{I}$ only appear in the first term of (\ref{Wsp}) as a shift in the thermodynamic grand potential per unit volume, so they can be ignored. The values of the real parts $\rho_{R}$ and $\phi_{R}$ still have to be determined by solving the saddle-point equations,%
\begin{equation}
\left\{
\begin{array}{l}%
\left.  \dfrac{\partial\Omega_{sp}\left(  \beta,\mu,\zeta;\rho_{R},\phi
_{R}\right)  }{\partial\rho_{R}}\right\vert _{\beta,\mu,\zeta;\phi_{R}}=0\\
\left.  \dfrac{\partial\Omega_{sp}\left(  \beta,\mu,\zeta;\rho_{R},\phi
_{R}\right)  }{\partial\phi_{R}}\right\vert _{\beta,\mu,\zeta;\rho_{R}}=0
\end{array}
\right.  \text{.} \label{saddlepointequations}%
\end{equation}
This way we obtain the extrema of $\Omega_{sp}\left(  \beta,\mu,\zeta;\rho_{R},\phi_{R}\right)$ as a function of $\rho_{R}$ and $\phi_{R}$.

Before continuing, it is interesting to look at the physics behind expression (\ref{Wsp}) for the thermodynamic grand potential per unit volume. The density fields have two effects on $\Omega_{sp}\left(  \beta,\mu,\zeta;\rho_{R},\phi_{R}\right)$. First, they result in an extra interaction energy, given by the second term of (\ref{Wsp}). Second, the third term of (\ref{Wsp}) corresponds to the thermodynamic grand potential per unit volume of the non-interacting two-component Fermi gas, except for the shifted chemical potentials. As such, the density fields can change the polarization of the gas. Note also that when half the density difference $\phi$ equals the average density $\rho$, something we would expect in the fully polarized case, only the chemical potential shift remains. This is the expected result for the fully polarized itinerant ferromagnetic state.

To ease our notation, from now on we will drop the index $R$ of $\rho_{R}$ and $\phi_{R}$. By performing the Matsubara summation in (\ref{Wsp}), we finally find%
\begin{align}
& \Omega_{sp}\left(  \beta,\mu,\zeta;\rho,\phi\right)=\frac{1}{g}\left(
\phi^{2}-\rho^{2}\right)\nonumber\\
&  -\int\frac{d^{D}\mathbf{k}}{\left(  2\pi\right)  ^{D}}\left\{  \frac
{1}{\beta}\ln\left[  \frac{1}{2}\cosh\left(  \beta\zeta^{\prime}\right)
+\frac{1}{2}\cosh\left(  \beta E_{\mathbf{k}}\right)  \right]  -E_{\mathbf{k}%
}\right\} \label{Wfinal}
\end{align}
in D dimensions. The integral over $k$ cannot be solved exactly for general cases (although it
can be solved exactly in some cases, for example for $D=2$ and in the zero temperature limit).

\section{The stability of the solution\label{stability}}

The saddle-point thermodynamic grand potential per unit volume $\Omega_{sp}\left(\beta,\mu,\zeta\right)$ can be calculated by solving the saddle-point equations (\ref{saddlepointequations}) for each value of the thermodynamic variables $\beta$, $\mu$ and $\zeta$. Here, we ask ourselves the following question: assuming that a certain set of parameter values $\left(  \beta,\mu,\zeta;\rho,\phi\right)$ is a solution to the saddle-point equations, would that solution be stable to quantum fluctuations? For the solution to be stable, it has to be a minimum of $\Omega_{sp}\left(  \beta,\mu,\zeta;\rho,\phi\right)$ as a function of $\rho$ and $\phi$ for a given set of thermodynamic variables $\left(  \beta,\mu,\zeta\right)$.

To study the nature of an extremum, we investigate the Hessian matrix of the saddle-point thermodynamic grand potential per unit volume $\Omega_{sp}\left(  \beta,\mu,\zeta;\rho,\phi\right)  $,
\begin{equation}
H=%
\begin{pmatrix}
\left.  \dfrac{\partial^{2}\Omega_{sp}\left(  \beta,\mu,\zeta;\rho,\phi\right)}{\partial\rho^{2}}\right\vert
_{\beta,\mu,\zeta;\phi} & \left.  \dfrac{\partial^{2}\Omega_{sp}\left(  \beta,\mu,\zeta;\rho,\phi\right)}{\partial
\rho\partial\phi}\right\vert _{\beta,\mu,\zeta}\\
\left.  \dfrac{\partial^{2}\Omega_{sp}\left(  \beta,\mu,\zeta;\rho,\phi\right)}{\partial\phi\partial\rho}\right\vert
_{\beta,\mu,\zeta} & \left.  \dfrac{\partial^{2}\Omega_{sp}\left(  \beta,\mu,\zeta;\rho,\phi\right)}{\partial\phi^{2}%
}\right\vert _{\beta,\mu,\zeta;\rho}%
\end{pmatrix}
\text{.} \label{hessian}%
\end{equation}
Calculating the second derivatives of (\ref{Wfinal}) in (\ref{hessian}), we find%
\begin{equation}
H=%
\begin{pmatrix}
-\frac{2}{g}-I_{1} & I_{2}\\
I_{2} & \frac{2}{g}-I_{1}%
\end{pmatrix}
\end{equation}
with $I_{1}$ and $I_{2}$ referring to the integrals%
\begin{equation}
I_{1}=\beta\int\frac{d^{D}\mathbf{k}}{\left(  2\pi\right)  ^{D}}\left(
\frac{1+\cosh\left(  \beta\zeta^{\prime}\right)  \cosh\left(  \beta
E_{\mathbf{k}}\right)  }{\left[  \cosh\left(  \beta\zeta^{\prime}\right)
+\cosh\left(  \beta E_{\mathbf{k}}\right)  \right]  ^{2}}\right)  \text{,}%
\end{equation}%
\begin{equation}
I_{2}=\beta\int\frac{d^{D}\mathbf{k}}{\left(  2\pi\right)  ^{D}}\left(
\frac{\sinh\left(  \beta E_{\mathbf{k}}\right)  \sinh\left(  \beta
\zeta^{\prime}\right)  }{\left[  \cosh\left(  \beta\zeta^{\prime}\right)
+\cosh\left(  \beta E_{\mathbf{k}}\right)  \right]  ^{2}}\right)  \text{.}%
\end{equation}

For an extremum to be a minimum of the thermodynamic grand potential per unit volume, the Hessian matrix has to have two positive eigenvalues. If at least one of its eigenvalues is negative, the solution is unstable. The trace of a matrix is equal to the sum of its eigenvalues. $\Tr H=-2I_{1}\leq0$ (because the hyperbolic cosine is a positive function), so at least one of the eigenvalues of $H$ is negative for all possible values of the parameters $g$, $\beta$, $\mu$, $\zeta$, $\rho$ and $\phi$. If this is true for all values, it is also true for the values that solve the saddle-point equations, implying all solutions are unstable to small (density) fluctuations.

\section{Conclusion and discussion\label{conclusiondiscussion}}

The main conclusion of this paper is that none of the extrema of the action for a homogeneous two-component Fermi gas with contact interactions are a minimum, if the Hubbard-Stratonovich transformation is performed with only density fields. This means that those extrema do not correspond to a (meta)stable state of the system, since they are unstable to small (quantum) fluctuations. Of course (stable) minima of the action itself do exist, but for contact interactions they cannot be found or approximated in the Hartree channel (using density fields). For low temperatures, a well-known minimum of the action is found in the Bogoliubov channel, which is used to describe superfluidity.

When considering density fields, small fluctuations are in fact density fluctuations. Consequently, we have studied the mechanical stability of the extrema: stability against collapse, expansion and spin separation (or a mix of these three). The instability we found indicates that the normal and polarized (itinerant ferromagnetic) states of a homogeneous two-component Fermi gas with contact interactions are mechanically unstable.

In experiments with ultracold atomic gases, we already know that the presence of the closed scattering channel causes an instability which prevents the formation of the itinerant ferromagnetic state (or more generally the formation of any equilibrium state in the short time scales of the experiment). In light of the proposals to circumvent this problem, it would be interesting to know if there are any other instabilities which would inhibit the experimental realization of the itinerant ferromagnetic state.

Our work can answer this question for the contact potential: in the path-integral formalism, the itinerant ferromagnetic state is unstable to density fluctuations. Since the details of the interaction potential may have a big influence on the stability of the system, further study is needed to determine if similar instabilities occur for other interaction potentials. We can only go beyond mean-field if we find a regime in which the extrema are stable (minima), so this stability analysis will be crucial for an improved theoretical description of itinerant ferromagnetism.

If we use the Hartree channel, we assume that the dominant interaction effects in the system are related to density fluctuations. That assumption fails when all saddle-points are unstable. In that case there are other interaction effects present which are more important than the density fluctuations (e.g. pair formation, leading to superfluidity).

What kind of potential would stabilize the itinerant ferromagnetic state? Would a short-ranged potential be sufficient, or is a long-ranged potential required? Using Tan's relations, it can be shown that the zero-range two-component Fermi gas does not have a stable fully polarized ferromagnetic state \cite{Tan1d}. As Tan's relations should be valid to a good approximation for finite-range interactions of sufficiently short range, this suggests that a long ranged interaction is needed to stabilize the itinerant ferromagnetic state. However, other work has predicted the occurrence of itinerant ferromagnetism for effective short-ranged interactions \cite{ferrolattice,ferropolarons}.

Zero-range potentials in Fermi gases are necessarily spin-dependent: between two equal spins the strength of a contact potential is always zero. This is no longer strictly true when the potential acquires a finite range, and (together with the fact that the kinetic energy is different) this may explain why Kollar \textit{et al.} \cite{ferrolattice} find that nearest neighbour interactions in a lattice model allow for itinerant ferromagnetism. On the other hand, long-range interactions can overcome spatial unmixing effects and increase domain sizes in the itinerant ferromagnetic state. So, our suggestion would be to look at long-range potentials for the Fermi gases, combined with a short-range repulsion to enhance the exchange effects.

A generalization of our approach to the case of a confining external potential can be made by using the Local Density Approximation (LDA). LDA relies on the assumption that the gas can be described by local thermodynamic variables (chemical potentials). Locally, the gas is approximated by the homogeneous gas at constant chemical potentials with the same values of the thermodynamic variables, which is the case we considered in the paper. Because we found an instability for \textit{all} values of the thermodynamic variables in a homogeneous gas with contact interactions, we can rule out itinerant ferromagnetism for a two-component Fermi gas with contact interactions in an external potential as long as LDA is valid.

Two requirements need to be fulfilled for LDA to be valid. First, the density (and therefore also the external potential) needs to be sufficiently slowly varying. This assumption is true for many, but not all confining external potentials used in experiments with ultracold atomic gases, and it generally breaks down at the edge of the atomic cloud. The second requirement is that the interactions can also be considered to be local, hence that the range of the interactions is sufficiently small. The second requirement is always fulfilled for contact interactions.

Summarizing, we find that stability is an important factor to consider when studying itinerant ferromagnetism. Furthermore, in the context of ultracold atomic gases we suggest to use a more realistic potential than the contact potential, preferably a long-ranged potential with a short-ranged repulsion.
\newline
\newline
\textbf{Acknowledgements} \textit{The authors would like to thank C.A.R. S\'{a} de Melo and A. Pelster for useful discussions, and J.P.A. Devreese for a careful reading of the manuscript. E. Vermeyen gratefully acknowledges support in the form of a Ph. D. fellowship of the Research Foundation - Flanders (FWO). This work was supported by the following Research Programmes of the Research Foundation-Flanders (FWO): G.0119.12.N, G.0115.12.N, G.0180.09.N, and G.0429.15.N. This work was also supported by the Research Council of Antwerp University via a GOA grant.}

\bibliographystyle{spphys}       
\bibliography{EVJT_IFstability}

\end{document}